\documentclass[12pt]{amsart}

\usepackage{amsmath,amssymb,amsfonts,amsthm,bbm}
\usepackage{graphicx}
\usepackage{float}
\usepackage{hyperref}
\usepackage{chngcntr}
\counterwithin{figure}{section}
\date{}
\usepackage{enumerate}
\usepackage{fullpage}

\newtheorem{theorem}{Theorem}[section]

\newtheorem{conjecture}[theorem]{Conjecture}

\theoremstyle{definition}

\newtheorem{remark}[theorem]{Remark}

\numberwithin{equation}{section}

\begin{document}

\title{\bf On the distribution of the maximum value of the characteristic polynomial of GUE random matrices}
\author{Y. V. Fyodorov and N. J. Simm }

\address{Queen Mary University of London, School of Mathematical Sciences, London E1 4NS, United Kingdom
\footnote{ \small e-mails: {\it y.fyodorov@qmul.ac.uk}  and {\it n.simm@qmul.ac.uk}}}

\begin{abstract}
Motivated by recently discovered relations between logarithmically correlated Gaussian processes and characteristic polynomials of large random $N \times N$ matrices $H$ from the Gaussian Unitary Ensemble (GUE), we consider the problem of characterising the distribution of the global maximum of $D_{N}(x):=-\log|\det(xI-H)|$ as $N \to \infty$ and $x\in (-1,1)$. We arrive at an explicit expression for the asymptotic probability density of the (appropriately shifted) maximum by combining the rigorous Fisher-Hartwig asymptotics due to Krasovsky \cite{K07} with the heuristic {\it freezing transition} scenario for logarithmically correlated processes. Although the general idea behind the method is the same as for the earlier considered case of the Circular Unitary Ensemble, the present GUE case poses new challenges. In particular we show how the conjectured  {\it self-duality} in the freezing scenario plays the crucial role in our selection of the form of the maximum distribution. Finally, we demonstrate a good agreement of the found probability density with the results of direct numerical simulations of the maxima of $D_{N}(x)$.

\end{abstract}
\maketitle
\section{Introduction.}
\begin{figure}
\includegraphics[width=400pt]{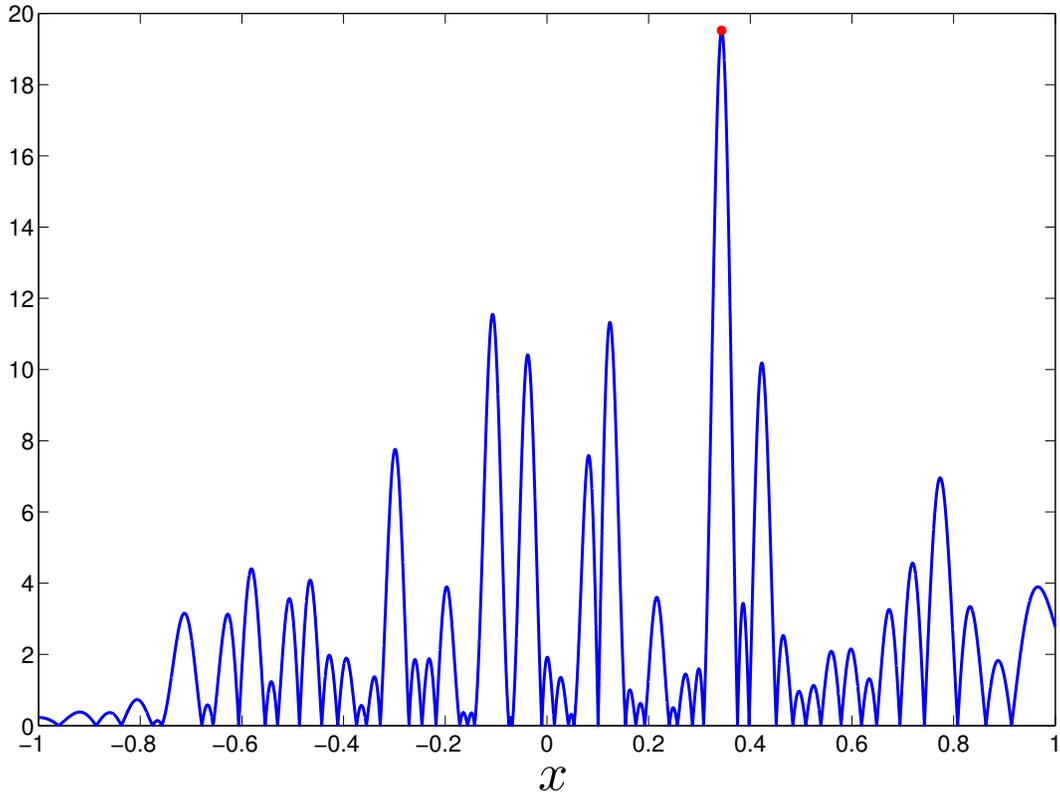}
\caption{A plot of a single realization of $|p_{N}(x)|e^{-\mathbb{E}\log|p_{N}(x)|}$ for $N=50$. The global maximum is marked with a red circle. }
\label{fig:guerealizationN50}
\end{figure}

\begin{figure}[H]
\includegraphics[width=400pt]{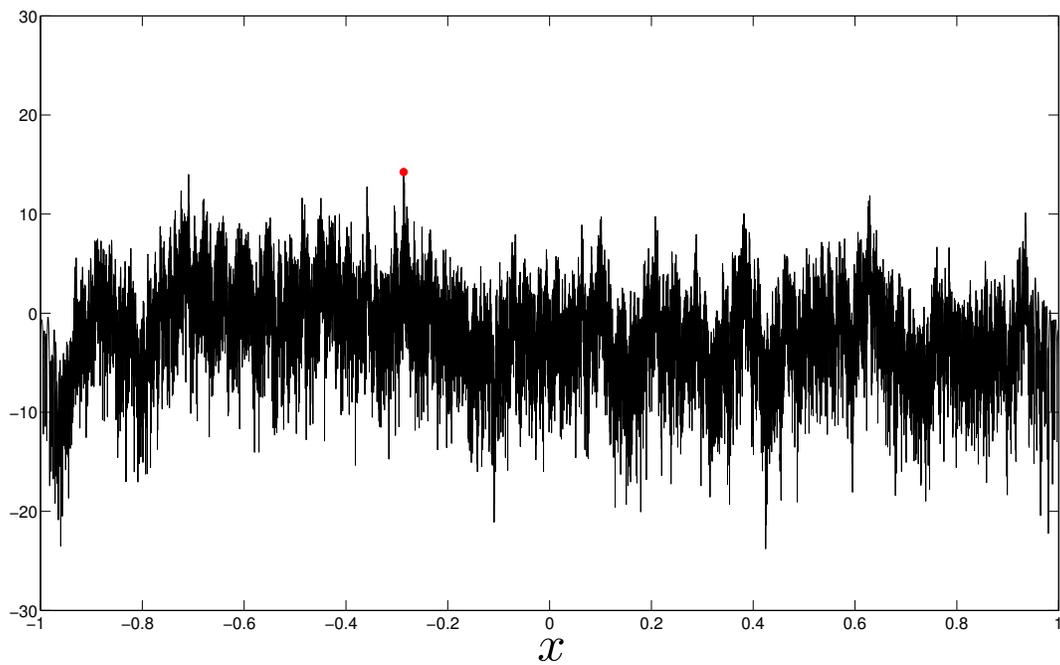}
\caption{A plot of  a single realization of $2\log \left(|p_{N}(x)
|e^{-\mathbb{E}\log|p_{N}(x)|}\right)$ with $N=3000$. The maximum value is marked with a red circle.}
\label{fig:N3000gueres}
\end{figure}

The space of all $N \times N$ Hermitian matrices $H$ with probability density function
\begin{equation}
\label{guedensity}
P(H) \propto \mathrm{exp}(-2N\mathrm{Tr}(H^{2}))
\end{equation}
is known as the \textit{Gaussian Unitary Ensemble} (or GUE)\cite{AGZ09,Meh04,PS11}. Here and henceforth the variance is chosen to ensure that asymptotically for $N\to \infty$, the limiting mean density of the GUE eigenvalues is given by the Wigner semicircle law $\rho(x) = (2/\pi)\sqrt{1-x^{2}}$ supported in the interval $x\in [-1,1]$. The characteristic polynomial $p_{N}(x) = \mathrm{det}(xI-H)$ of the matrix $H$ constitutes one of the most basic quantities of interest, encoding all eigenvalues of $H$ through the roots of $p_{N}(x)$.
 As one varies the argument $x$ over an interval containing many eigenvalues for a given realization of the ensemble, the value of the polynomial $p_N(x)$ shows huge variations by the orders of magnitude for large $N$, see Figure \ref{fig:guerealizationN50} for $N=50$ and Figure \ref{fig:N3000gueres}  for $N=3000$.

The purpose of this article is to describe the statistical properties of the \textit{highest peak} displayed by the modulus of the GUE polynomial $|p_{N}(x)|$, namely the probability density for the maximum value attained by $|p_{N}(x)|$ over the interval $[-1,1]$ on the real line as $N \to \infty$. Our main result is the following

\begin{conjecture}\label{pred}
Consider the random variable
\begin{equation}
M_{N}^{*} := \max_{x \in [-1,1]}\bigg\{2\log|p_{N}(x)|-2\mathbb{E}(\log|p_{N}(x)|)\bigg\} \label{maxrv}
\end{equation}
Then in the limit $N \to \infty$ we have
\begin{equation}
M_{N}^{*} = 2\log(N) - \frac{3}{2}\log(\log(N))-(1+o(1))y+o(1) \label{predeq}
\end{equation}
where $y$ is a continuous random variable characterized by the two-sided Laplace transform of its probability density:
\begin{equation}\label{Laplace}
\mathbb{E}(e^{ys}) = \frac{1}{C}(2\pi)^{s}\frac{\Gamma(s+1)\Gamma(s+3)G(s+7/2)^{2}}{G(s+6)G(s+1)}
\end{equation}
where $\Gamma(z)$ and $G(z)$ stand for the Euler gamma-function and the Barnes digamma-function, correspondingly. The normalization $C$ can be evaluated explicitly as
\begin{equation}
C = \frac{e^{1/4}\pi^{5/2}}{2^{9+11/12}A^{3}}
\end{equation}
where $A$ is the Glaisher-Kinkelin constant $A=e^{1/12-\zeta'(-1)} = 1.2824271291...$.
\end{conjecture}

\begin{remark}
The product form of the Laplace transform \eqref{Laplace} offers an interesting interpretation of the above results. Noting that $\Gamma(1+s)$ is the moment generating function of a standard Gumbel random variable $G$, we can write
\begin{equation}
y = G + y' \label{uprime}
\end{equation}
where $y'$ is an independent random variable with two-sided Laplace transform
\begin{equation}
\mathbb{E}(e^{y's}) = \frac{1}{C}(2\pi)^{s}\frac{\Gamma(s+3)G(s+7/2)^{2}}{G(s+6)G(s+1)}. \label{uprimelaplace}
\end{equation}
In the end of the paper we provide convincing numerical evidence that this Laplace transform does indeed define a unique random variable $y'$. This immediately implies that the probability density of $y$ is the convolution of a Gumbel random variable with $y'$. Such a convolution structure is expected to appear universally when studying the extreme value statistics of logarithmically correlated Gaussian fields, see the discussion around and after Eq. (\ref{LCdef}).
\end{remark}

In recent years, much interest has accumulated regarding the statistical behaviour of characteristic polynomials of various random matrices as a function of the spectral variable $x$. To a large extent this interest was stimulated by the established paradigm that many statistical properties of the Riemann zeta function along the critical line, that is $\zeta(1/2+it)$, can be understood by comparison with analogous properties of the characteristic polynomials of random matrices \cite{KS00,HKOC01,CFKRS05,CFZ,GHK,ABH}.

For invariant ensembles \cite{Meh04,PS11} of self-adjoint matrices with real eigenvalues, statistical characteristics of $p_{N}(x)$ depend very essentially on the choice of scale spanned by the real variable $x$. From that end it is conventional to say that $x$ spans the {\it local} (or {\it microscopic}) scale if one considers intervals containing in the limit $N\to \infty$ typically only a finite number of eigenvalues (the corresponding scale for GUE in (\ref{guedensity}) is of the order of $1/N$). At such scales, standard objects of interest are correlation functions containing products and ratios of characteristic polynomials, which show determinantal/Pfaffian  structures \cite{BH2000,FS03a,FS03b,BDS,BS06,KG10} for Hermitian/real symmetric matrices and tend to universal limits at the local scale. Similar structures arise for properly defined characteristic polynomials $p_N(\theta)=\det{\left(I-U\,e^{-i\theta}\right)}$ of circular ensembles
 (like CUE, COE, and CSE)\cite{Meh04} of unitary random matrices $U$ uniformly distributed with respect to the Haar measure on $U(N)$ (and other classical groups) \cite{CFKRS05,BG06,CFS}, whose properties on the local scale are indistinguishable from their Hermitian counterparts.

Next, when $x$ spans an interval containing in the limit $N\to \infty$ typically of order of $N$ eigenvalues one speaks of the {\it global } (or {\it macroscopic}) scale behaviour. At such a scale properties of $p_N(x)$ display both universal and non-universal features, the latter depending on the ensemble chosen. The study of characteristic polynomials at such a scale was initiated in \cite{HKOC01} where it was shown that the function $V_N(\theta)=-2\log{|\det(1-U\,e^{-i\theta})|}$, with $U$ belonging to the CUE, converges (in an appropriate sense) to a random Gaussian Fourier series of the form
\begin{equation} \label{1/f}
 V(\theta)=\sum_{n=1}^{\infty}\frac{1}{\sqrt{n}} \left(v_n e^{i n \theta}+\overline{v_n} e^{-i n \theta}\right)\,,
\end{equation}
where the coefficients $v_n,\overline{v}_n$ are independent standard complex Gaussian random variables, i.e. $\mathbb{E} \{v_n\}=0$, $\mathbb{E}\{v_{n}^{2}\}=0$ and $\mathbb{E} \{v_n \overline{v}_n \}=1$. The covariance structure associated with such a process is given by $\mathbb{E} \{V(\theta_1)V(\theta_2)\}=-2\log {|e^{i\theta_1}-e^{i\theta_2}|}$ as long as $\theta_1\ne \theta_2$. Such a (generalized) random function $V(\theta)$ is a representative of random processes known in the literature under the name of {\it 1/f noises},
see \cite{FLDR12,FK14} for background discussion and further references.

Recently the study of the global scale behaviour was extended to the GUE polynomial $p_N(x)$ in \cite{FKS13} by using earlier insights from \cite{Joh98} and \cite{K07}. That work revealed again a structure analogous to that of (\ref{1/f}), though different in detail. Namely, it was shown that the natural limit of $\tilde D_N(x):=-\log |p_N(x)| + \mathbb{E}\{ \log |p_N(x)| \}$ is given by the random Chebyshev-Fourier series
\begin{equation}\label{1/fch}
F(x) = \sum_{n=1}^{\infty}\frac{1}{\sqrt{n}}\, a_{n}\, T_{n}(x), \qquad x \in (-1,1),
\end{equation}
 with $T_{n}(x) = \cos(n\arccos(x))$ being Chebyshev polynomials and real $a_n$ being independent standard Gaussians.  A quick computation shows that the covariance structure associated with the generalized process $F(x)$ is given by an integral operator with kernel
\begin{equation}\label{covop}
 \mathbb{E}\{F(x)F(y)\} = \sum_{n=1}^{\infty}\frac{1}{n}T_{n}(x)T_{n}(y) = -\frac{1}{2}\log(2|x-y|),
\end{equation}
 as long as $x\ne y$. Such a limiting process $F(x)$ is an example of an aperiodic $1/f$-noise.

Finally, one can consider an intermediate, or {\it mesoscopic} spectral scales, with intervals typically containing in the limit $N\to \infty$ the number of eigenvalues growing with $N$, but representing still a vanishingly small fraction of the total number $N$ of all eigenvalues. The properties of the characteristic polynomials at such scales were again addressed in \cite{FKS13} where it was shown that for the GUE, that object gives rise to a particular (singular) instance of the so-called fractional Brownian motion (fBm) \cite{ManvNess68,Taq2003} with the Hurst index $H=0$, again characterized by correlations logarithmic in the spectral parameter.

The discussion above serves, in particular, the purpose of pointing to an intimate connection between Gaussian random processes with logarithmic correlations and the modulus of characteristic polynomials at global and mesoscopic scales. The relation is important as logarithmically correlated  Gaussian (LCG) random processes and fields attract growing attention in Mathematical Physics and Probability and play an important role in problems of Quantum Gravity,  Turbulence, and Financial Mathematics, see e.g. \cite{DRSV14}. In particular, the periodic $1/f$ noise (\ref{1/f}) emerged in constructions of conformally invariant planar random curves \cite{AJKS11}. Among other things, the statistics of the global maximum of LCG fields attracted considerable attention, see  \cite{DRZ2015} and references therein. Particularly relevant in the present context are the results of Ding, Roy and Zeitouni \cite{DRZ2015} on the maxima of regularized lattice versions of LCG fields which we discuss informally below. Let $V_N = \mathbb{Z}_N^d$
be the $d-$dimensional box of side length $N$ with the left bottom corner located at the origin.
 A suitably normalized version of the logarithmically correlated Gaussian field is a collection of Gaussian variables {$\phi_{N,v} : v \in V_N$} with variance $\mathbb{E}\{\phi^2_{N,v}\}=2\log{N}+ f(v)$ and covariance structure
\begin{equation}\label{LCdef}
\mathbb{E}\{\phi_{N,v},\phi_{N,u}\}= 2\log_{+}\frac{N}{|u - v|} + g(u,v), \quad \mbox{ for}\,\,  \,\, u \ne v \in V_N
\end{equation}
where $\ln_+(w) = \max\left(\ln{w}, 0\right)$ and both $f(v)$ and $g(u,v)$  are continuous bounded functions far enough from the boundary of $V_N$. Now set $M_N =\max_{v\in V_N}\phi_{N,v}$ and $m_N = \sqrt{d} \log{N} - \frac{3}{2d}\log\log{N}$. The limiting law of $M_N-m_N$ is then expected, after an appropriate shift and rescaling, to be given by the {\it Gumbel distribution with random shift}:
\begin{equation}\label{LCmax}
P(y)=\lim_{N\to \infty} Prob(M_N \ge m_N-y)=\mathbb{E}\left\{e^{-e^{\sqrt{d}(y-z)}}\right\},
\end{equation}
where the distribution of the random shift variable $z$ depends on details
of the behaviour of covariance (\ref{LCdef}) for $|u - v|\sim N$ and $|u - v|\sim 1$, see the detailed discussion in  \cite{DRZ2015}. The random variable $Z=e^{-\sqrt{d}z}$ is related to the so called {\it derivative martingale} associated with the LCG fields \cite{DRZ2015} whose distribution is however not known. Recently it has been shown that the recentering term $m_{N}$ in \eqref{LCmax} also holds for a randomized model of the Riemann zeta function \cite{ABH}, proved by revealing a special branching structure within the associated logarithmic correlations.

 We see that our conjecture \ref{pred} for the maximum of characteristic polynomial of large GUE matrices fully agrees with the predicted structure of the maximum of LCG in dimension $d=1$.  Note that the expression (\ref{LCmax}) implies that the double-sided Laplace transform  of the density $\rho(y)=-\frac{d}{dy} P(y)$ for the (shifted) maximum $y$ is related to the density $\tilde{\rho}(z)$ of the random variable $z$ as
\begin{equation}
\mathbb{E}\left(e^{ys}\right)=\int \rho(y) e^{sy}\,dy=\Gamma(s+1) \int \tilde{\rho}(z) e^{sz}\,dz=\Gamma(s+1) \mathbb{E}\left(e^{zs}\right)
\end{equation}
which is in turn equivalent to the Gumbel convolution in eq.(\ref{uprime}). In fact our formula (\ref{uprimelaplace}) provides the explicit form of the distribution for the derivative martingale of our model, thus going considerably beyond the considerations of \cite{DRZ2015}.

From a quite different perspective, processes similar to (\ref{1/f}) and (\ref{1/fch}) appeared in the context of statistical mechanics of disordered systems when studying extreme values of random multifractal landscapes supporting spinglass-like thermodynamics \cite{FB08,FLDR09,FLDR12,AZ14}. The latter link is especially important in the context of the present paper. The idea that it is beneficial to look at $|p_{N}(\theta)|$ as a disordered landscape consisting of many peaks and dips, and to think of an associated statistical mechanics problem was put forward in \cite{FHK12,FK14}. It allowed to get quite non-trivial analytical insights into statistics of the maximal value of the CUE polynomial sampled over the full circle $\theta\in [0,2\pi]$, or over its mesoscopic sub-intervals. This was further used to conjecture the associated properties of the modulus of the Riemann zeta-function along the critical line, see
some recent advances inpired by that line of research in \cite{ABH}. Some relations between between CUE characteristic polynomials and logarithmically correlated processes (in the form of the so-called "multiplicative chaos" measures introduced by Kahane, see \cite{RV2014} for a review) was recently rigorously verified in \cite{Webb14}. The case of GUE polynomials however remained outstanding.

It is our objective in this paper to provide two separate means of supporting Conjecture \ref{pred}. First, we will provide careful and explicit, albeit in part heuristic, analytical arguments. Although our technique is inspired by the approach of \cite{FK14} it contains new nontrivial features necessary to overcome challenges arising from the non-uniform eigenvalue density $\rho(x)$, reflecting absence of translational invariance for the GUE at the global spectral scale (note e.g. the non-trivial recentering in \eqref{maxrv}). All this makes actual calculation for the GUE much more involved in comparison to the CUE and the limiting random variable $u$ above appears to be more complicated than its CUE counterpart. Secondly, we will test our Conjecture with numerical experiments for matrices of size $N=3000$ and around $250,000$ realizations. This is especially important as part of our analysis is based on very plausible but as yet not fully rigorous considerations.
Finally, is natural to expect that the same distribution should be shared by the maximum modulus of characteristic polynomials for Hermitian random matrices with independent entries taken from the so-called Wigner ensembles, see  \cite{EYY12}.

Before giving the detail of our procedure in the next section we need to quote the following fundamental asymptotic result obtained by Krasovsky \cite{K07} which will be central for our considerations:
\begin{align}
\mathbb{E}\left(\prod_{j=1}^{k}|p_{N}(x_{j})|^{2\alpha_{j}}\right) &= \prod_{j=1}^{k}C(\alpha_{j})(1-x_{j}^{2})^{\alpha_{j}^{2}/2}(N/2)^{\alpha_{j}^{2}}e^{(2x_{j}^{2}-1-2\log(2))\alpha_{j}N}\\
&\times \prod_{1 \leq i < j \leq k}(2|x_{i}-x_{j}|)^{-2\alpha_{i}\alpha_{j}}\left[1+O\left(\frac{\log N}{N}\right)\right] \label{krasov}
\end{align}
where
\begin{equation}
C(\alpha) := 2^{2\alpha^{2}}\frac{G(\alpha+1)^{2}}{G(2\alpha+1)}
\end{equation}
and $G(z)$ is the Barnes G-function. Differentiating with respect to $\alpha$, we deduce that
\begin{equation}
\mathbb{E}(2\log|p_{N}(x)|) = N(2x^{2}-1-2\log(2)) + O(\log(N)/N),
\end{equation}
where we used that $C'(0)=0$.

The most salient feature of the asymptotics \eqref{krasov} is the product of differences on the second line, which when rewritten in the form
\begin{equation}
\exp \Big[ -\sum_{1 \leq i < j \leq k}2\alpha_{i}\alpha_{j}\,\log|2(x_{i}-x_{j})|\Big] , \label{kracovstruct}
\end{equation}
 can be looked at as evidence of the limiting Gaussian process (\ref{1/fch}) with logarithmic covariance (\ref{covop}) in the background. We will however stress that naively replacing the (shifted) $\log|p_{N}(x_{j})|$ with the corresponding $1/f$ noise (\ref{1/fch})  is not a valid approximation as the factors $(1-x_{j}^{2})^{\alpha_{j}^{2}/2}$ in (\ref{krasov}) do play an essential role in determining the extreme value statistics of $|p_{N}(x_{j})|$. Let us finally note that had we suppressed the factors $C(\alpha_{j})$ the faithful description of $\log|p_{N}(x_{j})|$ would be that of the regularized LCG process with covariance (\ref{covop}), the position-dependent variance $2\ln{N}+2\ln{\frac{1}{2}\sqrt{1-x^2}}$ and the position-dependent mean $N(2x^2-1-2\log(2))$.

\textbf{Acknowledgements:} We are grateful for helpful comments from Christian Webb during the preparation of this manuscript. We acknowledge support from EPSRC grant EP/J002763/1 ``\textsf{Insights into Disordered Landscapes via Random Matrix Theory and Statistical Mechanics}''. \\\\

\section{Statistical mechanics approach to the distribution of GUE characteristic polynomials.}

 Following the ideas of \cite{FK14} we recast the problem of computing the value of the global maximum of $|p_{N}(x)|$ (with an appropriate shift by the mean value) as a statistical mechanics problem characterized by the partition function
\begin{equation}\label{part}
\mathcal{Z}_{N}(\beta) = \frac{N}{2}\int_{-1}^{1}e^{-\beta \phi_{N}(x)}\,\rho(x)^{q}\,dx, \qquad \beta>0 , \, q\ge 0
\end{equation}
with the "potential" $\phi_{N}(x) = -2\left(\log|p_{N}(x)|-\mathbb{E}\log|p_{N}(x)|\right))$, inverse temperature $\beta>0$ and $\beta$-independent non-negative parameter $q$.
Specifically, if we define the associated "free energy" as $\mathcal{F}(\beta)=-\beta^{-1}\log{\mathcal
{Z}_N(\beta)}$, then
\begin{equation}\label{freeen}
\lim_{\beta\to \infty}\mathcal{F}(\beta)=
\min_{x\in(-1,1)}\phi_N(x)=2\max_{x\in(-1,1)}\left[\log{|p_N(x)|-\mathbb{E}\log|p_{N}(x)|}\right].
\end{equation}

Note that if compared to a similar partition function for the CUE case the main new feature in \eqref{part} is the factor $\rho(x)^q$. Although naively the presence of such a factor may seem irrelevant when taking the limit $\beta \to \infty$, we will actually see that it plays a very important role in supporting our procedure of extracting the free energy for $\beta$ exceeding some critical value.

Now we aim to compute the integer moments of the partition function defined in \eqref{part}:
\begin{equation} \label{mom}
\mathbb{E}(\mathcal{Z}^{k}_{N}(\beta)) = \left(\frac{N}{2}\right)^{k}\int_{-1}^{1}\ldots \int_{-1}^{1}\mathbb{E}\left( \prod_{j=1}^{k}|p_{N}(x_{j})|^{2\beta}\right)\prod_{j=1}^{k}e^{-2\beta\mathbb{E}\log|p_{N}(x_{j})|}\rho^{q}(x_{j})\,dx_{j}
\end{equation}
In the limit $N\to \infty$ the leading asymptotics of the above integral can be extracted by replacing the factor
$\mathbb{E}\left( \prod_{j=1}^{k}|p_{N}(x_{j})|^{2\beta}\right)$ with its asymptotics from \eqref{krasov}. In this way one obtains
\begin{align}\label{mompart}
\mathbb{E}(\mathcal{Z}^{k}_{N}(\beta)) \sim
 \left(\left(\frac{N}{2}\right)^{1+\beta^{2}}C(\beta)(2/\pi)^{q}\right)^{k}
 \int_{[-1,1]^{k}}\prod_{j=1}^{k}(1-x_{j}^{2})^{\frac{\beta^{2}+q}{2}}\prod_{1 \leq i < j \leq k}(2|x_{j}-x_{i}|)^{-2\beta^{2}}\,dx_{1}\ldots dx_{k}
\end{align}
After changing variables $x_{j}=2y_{j}-1$ the integral above assumes the form
\begin{align}\label{momselb} 2^{k(\beta^{2}+q+1)-2\beta^{2}k(k-1)}\int_{[0,1]^{k}}\prod_{j=1}^{k}y_{j}^{\frac{\beta^{2}+q}{2}}(1-y_{j})^{\frac{\beta^{2}+q}{2}}\prod_{1 \leq i < j \leq k}|y_{i}-y_{j}|^{-2\beta^{2}}\,dy_{1}\ldots dy_{k}
\end{align}
\[
= 2^{k(\beta^{2}+q+1)-2\beta^{2}k(k-1)}S_{k}\left(\frac{\beta^{2}+q}{2},\frac{\beta^{2}+q}{2},-\beta^{2}\right)
\]
with the quantity $S_{k}(a,b,-\gamma)$ being the well-known Selberg integral \cite{ForWar}:
\begin{align}\label{selb}
S_{k}(a,b,-\gamma) &:= \int_{[0,1]^{k}}\prod_{j=1}^{m}x_{j}^{a}(1-x_{j})^{b}\prod_{1 \leq i < j \leq k}|x_{i}-x_{j}|^{-2\gamma}\,dx_{1}\ldots dx_{k}\\
&= \prod_{j=1}^{k}\frac{\Gamma(a+1-(j-1)\gamma)\Gamma(b+1-(j-1)\gamma)\Gamma(1-j\gamma)}{\Gamma(a+b+2-(k+j-2)\gamma)\Gamma(1-\gamma)}\\
&=\frac{1}{\Gamma^{k}(1-\gamma)}\tilde{S}_{k}(a,b,-\gamma)
\end{align}

It is easy to see that the found expression for the partition function moments $\mathbb{E}(\mathcal{Z}_{\beta}^{k})$  in (\ref{momselb},\ref{mompart}) is well-defined for  any $0<\gamma=\beta^2<1$ and for an {\it integer} $k$ satisfying $1<k<\gamma^{-1}$.  To understand how to deal with the case $k>\frac{1}{\beta^2}$, we recall that Krasovsky's  asymptotic formula (\ref{krasov})
 is valid only when all of the differences $|x_i-x_j|$ remain {\it finite} when $N\to \infty$, and should be replaced by a different expression when  $ |x_i-x_j|\sim N^{-1}$. One can check that the divergence of the integral for $k>1/\beta^2$ is due precisely to the fact
 that these near degeneracies become important. Relying on our experience with the corresponding situation for the CUE \cite{FK14} case suggests that taking into account the correct short-scale cutoff  cures the formal divergence, but changes the asymptotics of the moments $\mathbb{E}(\mathcal{Z}_{\beta}^{k})$ with $N$: namely, these become of the order of $N^{1+k^2\beta^2}$ for
 $k>\beta^{-2}$ whereas they are of the order of $N^{(1+\beta^2)k}$ for $k<\beta^{-2}$.  Such a change of behaviour will lead to a log-normal (far) tail in the distribution. Note that for the CUE case, the above behaviour
 conjectured in \cite{FK14} was validated by recent rigorous calculation \cite{CK14}. There is no doubt
 that the same mechanism is operational in present case as well and will be validated by extending the theory
  of \cite{CK14} from Toeplitz to Hankel case. Actually, as argued in \cite{FB08} the
 moments with $k>\beta^{-2}$ play only a secondary role when addressing the question of extreme value statistics which is controlled exclusively by moments with $1<k<\beta^{-2}$. Our next goal is to use the latter integer moments for restoring the associated part of the probability density $\mathcal{P}(\mathcal{Z}_{\beta})$ for the partition function. This will be achieved if we manage to find the distribution for a random variable $z_{\beta}$ whose positive integer moments are given by
\begin{align}\label{zreduced}
\mathbb{E}(z_{\beta}^{k})=\tilde{S}_{k}(a,b,-\gamma), \quad a=b=\frac{q+\beta^2}{2}, \, \, \gamma=\beta^2
\end{align}
Such a task actually requires finding a way to continue analytically those moments to \textit{complex} $k$.
  Below we will arrive at the required continuation by exploiting a relatively simple heuristic procedure suggested in  \cite{FLDR09}.  Note that in a series of insightful papers \cite{Ost09,Ost12,Ost12a,Ost14} Ostrovsky developed a rigorous mathematical procedure of the required continuation which provides an {\it aposteriori}   justification of the results obtained via the heuristic approach.

\subsection{Analytical continuation of Selberg's Integral}

One starts with finding a recursion satisfied by $\tilde{S}_{k}(a,b,\gamma)$ for integer $k$ which is suitable for the continuation. By writing
\begin{equation}
\prod_{j=1}^{k-1}\Gamma(a+b+2-(k+j-3)\gamma) = \frac{\Gamma(2+a+b-(k-2)\gamma)}{\Gamma(2+a+b-(2k-2)\gamma)}\frac{\prod_{j=1}^{k}\Gamma(2+a+b-(k+j-2)\gamma)}{\Gamma(2+a+b-(2k-3)\gamma)}
\end{equation}
one sees immediately that
\begin{equation}\label{rec1}
\frac{\mathbb{E}(z_{\beta}^{k})}{\mathbb{E}(z_{\beta}^{k-1})} = \frac{\Gamma(a+1-(k-1)\gamma)\Gamma(b+1-(k-1)\gamma)\Gamma(1-k\gamma)\Gamma(2+a+b-(k-2)\gamma)}{\Gamma(2+a+b-(2k-2)\gamma)\Gamma(2+a+b-(2k-3)\gamma)}
\end{equation}
  It is convenient to introduce the moments $M_{\beta}(s)$ of the random variable $z_\beta$  defined for any complex $s$ as $M_{\beta}(s)=\mathbb{E}(z_{\beta}^{1-s})$ . We then have $\mathbb{E}(z_{\beta}^{k})=M(1-k),\, \mathbb{E}(z_{\beta}^{k-1})=M(2-k)$ and after identifying $s=1-k$ the recursion (\ref{rec1}) takes the form
\begin{equation}\label{rec2}
\frac{M_{\beta}(s)}{M_{\beta}(s+1)} = \frac{\Gamma(1+a+\gamma s)\Gamma(1+b+\gamma s)\Gamma(1+(s-1)\gamma)\Gamma(2+a+b+(s+1)\gamma)}{\Gamma(2+a+b+2s\gamma)\Gamma(2+a+b+(2s+1)\gamma)}
\end{equation}
which is now assumed to be valid for any complex $s$.
It is convenient to further use the duplication formula for the Gamma function :
\begin{equation}
\Gamma(2z) = \frac{2^{2z-1}}{\sqrt{\pi}}\Gamma(z)\Gamma(z+1/2)
\end{equation}
to get rid of the argument $2s$ in the denominator. Indeed, we have
\begin{align*}
&\Gamma(2+a+b+2\gamma s) = 2^{1+a+b+\gamma 2s}\Gamma(1+\gamma s + (a+b)/2)\Gamma(\gamma s+(a+b+3)/2)/\sqrt{\pi}\\
&\Gamma(2+a+b+\gamma(2s+1)) \\
&= 2^{1+a+b+\gamma(2s+1)}\Gamma(1+\gamma(s+1/2)+(a+b)/2)\Gamma(\gamma(s+1/2)+(a+b+3)/2)/\sqrt{\pi}
\end{align*}
so that (\ref{rec2}) assumes the form
\begin{align}\label{rec3}
\frac{M_{\beta}(s)}{M_{\beta}(s+1)} = &\frac{\Gamma(1+a+\gamma s)\Gamma(1+b+\gamma s)\Gamma(1+\gamma(s-1))\Gamma(2+a+b+(s+1)\gamma)}{\Gamma(1+\gamma(s+1/2)+(a+b)/2)\Gamma(\gamma(s+1/2)+(a+b+3)/2)}\\
& \times \frac{\pi}{2^{2(1+a+b)+(4s+1)\gamma}}\frac{1}{\Gamma(1+(a+b)/2+\gamma s)\Gamma((3+a+b)/2+\gamma s)}\nonumber
\end{align}
Recalling that according to (\ref{zreduced}) in our particular case $a=b=\frac{q}{2}+\frac{\beta^2}{2}$ we now use the parameterisation  $a=a_{1}+a_{2}\beta^{2}$, $b=b_{1}+b_{2}\beta^{2}$ and $\beta$-independent constants $a_1,a_2,b_1.b_2$. After this we finally arrive at
\begin{align}
&\frac{M_{\beta}(s)}{M_{\beta}(s+1)} = \\ \label{rec4}
&\frac{\Gamma(1+a_{1}+\beta^{2} (s+a_{2}))\Gamma(1+b_{1}+\beta^{2} (s+b_{2}))\Gamma(1+\beta^{2}(s-1))\Gamma(2+a_{1}+b_{1}+(s+1+a_{2}+b_{2})\beta^{2})}
{\Gamma(1+\beta^{2}(s+1/2+a_{2}/2+b_{2}/2)+(a_{1}+b_{1})/2)
\Gamma(\beta^{2}(s+(1+a_{2}+b_{2})/2)+(a_{1}+b_{1}+3)/2)}\\
& \times \frac{\pi 2^{-2(1+a_{1}+b_{1})-(4s+1+2a_{2}+2b_{2})\beta^{2}}}{\Gamma(1+(a_{1}+b_{1})/2+\beta^{2}(s+(a_{2}+b_{2})/2))
\Gamma((3+a_{1}+b_{1})/2+\beta^{2}(s+(a_{2}+b_{2})/2))}\nonumber
\end{align}
To determine the function $M_{\beta}(s)$ which satisfies (\ref{rec4}) for any complex $s$ we follow \cite{FLDR09} and introduce a variant of the Barnes function $G_\beta(x)$ which for any $\Re(x)>0$ is defined by:
\begin{align} \nonumber
&&   \ln G_\beta(x) = \frac{x-Q/2}{2} \ln (2 \pi) + \int_0^\infty \frac{dt}{t} \left( \frac{e^{- \frac{Q}{2} t} - e^{- x t}}{(1-e^{-\beta t})(1-e^{-t/\beta})}
+\frac{e^{-t}}{2} (Q/2-x)^2 + \frac{Q/2-x}{t} \right)
\end{align}
where $Q=\beta+1/\beta$. This function satisfies the so-called self-duality relation
\begin{eqnarray}\label{ZamoBarndual}
&& G_\beta(x) = G_{1/\beta}(x)
\end{eqnarray}
and further posesses a shift property that is central for our studies
\begin{eqnarray} \label{Gt1}
&&G_\beta(x + \beta) = \beta^{1/2 - \beta x}(2 \pi)^{\frac{\beta-1}{2}} \Gamma(\beta x)\,G_\beta(x)
\end{eqnarray}
One can check that $G_\beta(x)$ for $\beta=1$ coincides with the standard Barnes function $G(x)$ which is a unique solution of the recursion $G(x + 1)=\Gamma(x)G(x)$ satisfying $G(1)=1$. Similarly to the standard Barnes function the general Barnes $G_\beta(x)$ has no poles and only zeroes located at $x=-n \beta  - m/\beta$, $n,m=0,1,..$.
A detailed discussion of properties of functions closely related to  $G_\beta(x)$ can be found in \cite{Ost12,Ost12a}.

 Let us now define a function $M^{(G)}_{\beta}(s)$ of the complex argument $s$ by
\begin{align}\nonumber
&M^{(G)}_{\beta}(s) = \pi^{-s}2^{B_{1}s^{2}+B_{2}s}\beta^{\beta^{2}s}\\ \label{momG}
&\times\frac{G_{\beta}\left(\beta\left(s+\frac{a_{2}+b_{2}+1}{2}\right)+\frac{2+a_{1}+b_{1}}{2\beta}\right)
G_{\beta}\left(\beta\left(s+\frac{a_{2}+b_{2}}{2}\right)+\frac{2+a_{1}+b_{1}}{2\beta}\right)}
{G_{\beta}\left(\beta(s+a_{2})+\frac{1+a_{1}}{\beta}\right)G_{\beta}\left(\beta(s+b_{2})+\frac{1+b_{1}}{\beta}\right)}\\ \nonumber
&\times \frac{G_{\beta}\left(\beta\left(s+\frac{a_{2}+b_{2}+1}{2}\right)+\frac{3+a_{1}+b_{1}}{2\beta}\right)
G_{\beta}\left(\beta\left(s+\frac{a_{2}+b_{2}}{2}\right)+\frac{3+a_{1}+b_{1}}{2\beta}\right)}
{G_{\beta}\left(\beta(s+1+a_{2}+b_{2})+\frac{2+a_{1}+b_{1}}{\beta}\right)G_{\beta}\left(\beta(s-1)+\frac{1}{\beta}\right)}
\end{align}
where $B_{1} = 2\beta^{2}$ and $B_{2} = 2(a_{1}+b_{1}+1)+\beta^{2}(2a_{2}+2b_{2}-1)$. Then a straightforward computation which relies on the identity following from (\ref{Gt1})
\begin{equation}
\frac{G_{\beta}(\beta(s+1)+c/\beta)}{G_{\beta}(\beta s+c/\beta)}=(2\pi)^{\frac{\beta-1}{2}}\beta^{1/2-\beta^{2}s-c}\Gamma(c+\beta^{2}s)
\end{equation}
 shows that the ratio $\frac{M^{(G)}_{\beta}(s)}{M^{(G)}_{\beta}(s+1)}$ reproduces the right-hand side of (\ref{rec4})
 from which we conclude
\begin{equation}
\frac{M^{(G)}_{\beta}(s)}{M^{(G)}_{\beta}(s+1)} = \frac{M_{\beta}(s)}{M_{\beta}(s+1)}
\end{equation}
which finally implies that
\begin{equation}\label{compmom}
M_{\beta}(s) = M^{G}_{\beta}(s)\frac{M_{\beta}(1)}{M^{(G)}_{\beta}(1)}
\end{equation}
where $M_{\beta}(1) \equiv 1$. Together with \eqref{mompart}, \eqref{momselb} and the fact that $M_{\beta}(1)=1$, we obtain for $\beta<1$:
\begin{equation}
\label{znmoms}
\mathbb{E}(\mathcal{Z}_{N}(\beta)^{1-s}) \sim \left(\frac{\left(\frac{N}{2}\right)^{1+\beta^{2}}C(\beta)(2/\pi)^{q}}{\Gamma(1-\beta^{2})}\right)^{1-s}2^{(1-s)(\beta^{2}+q+1)+2\beta^{2}s(1-s)}\frac{M^{(G)}_{\beta}(s)}{M^{(G)}_{\beta}(1)}
\end{equation}

\subsection{Duality and the freezing transition}

The pair (\ref{momG})-(\ref{compmom}) solves the problem of finding the complex moments $M_{\beta}(s) =\mathbb{E}(z_{\beta}^{1-s})$ of the random variable
$z_{\beta}$ for any complex $s$, and $\beta<1$. Knowledge of such moments can be used to restore the probability distribution of $z_{\beta}$, hence of the partition function $\mathcal{Z}_{N}(\beta)$, and of its logarithm (the free energy) for large $N\gg 1$. Our goal is however to study the limit of the latter as $\beta\to \infty$ and one therefore should have a way of extracting information on the distribution for $\beta>1$. In doing this we rely on the {\it freezing transition scenario} for logarithmically correlated random landscapes. The background idea of such scenario goes back to \cite{CLD01} and was further advanced and clarified in the series of works \cite{FB08,FLDR09,FLDR10,FLDR12}. In brief, this scenario predicts a phase transition at the critical value $\beta=1$ and amounts to the following principle:
 \begin{center}
 {\it Thermodynamic quantities which for $\beta<1$ are {\bf self-dual} functions of the inverse temperature $\beta$, i.e. functions that remain invariant under the transformation $\beta\to \beta^{-1}$,
 retain for all $\beta>1$  the value they acquired at the point of self-duality $\beta =1$.}
 \end{center}
  Although such a scenario is not yet proven mathematically in full generality and has the status of a conjecture supported by physical arguments and available numerics, recently a few nontrivial aspects of freezing were verified within rigorous probabilistic analysis, see e.g. \cite{AZ14,DRZ2015,SZ2014} for efforts in this direction.

Within that scenario, one of the main outcomes of the analysis performed in  \cite{FLDR09} is that the self-dual object associated with the distribution of the partition function for logarithmically correlated
landscapes  is expected to be the appropriately defined Laplace transform:
 \begin{align}\label{main}
 g_\beta(y) = \mathbb{E}\left(\exp\left[ - e^{\beta y} \mathcal{Z}_{N}(\beta)/\mathcal{Z}^{e}_{N}(\beta)\right]\right),
\end{align}
where $\mathcal{Z}^{e}_{N}(\beta)$ is a typical scale of the partition function which is extracted from the asymptotic for the integer moments and in our case can be chosen as
\begin{equation}
\mathcal{Z}^{(e)}_{N}(\beta)= N^{1+\beta^{2}}\frac{\left[G(\beta+1)\right]^{2}}{G(2\beta+1)\Gamma(1-\beta^{2})}\left(\frac{4}{\pi}\right)^{q}.
\end{equation}
Moreover, defining the probability
density $p_{\beta}(y)$ by $p_{\beta}(y)=-g_{\beta}'(y)$ one can show that the double-sided Laplace transform for such a probability density is related to the complex moments $\tilde{M}_{\beta}(s) = \mathbb{E}\left(\frac{\mathcal{Z}_{N}(\beta)}{\mathcal{Z}^{(e)}_{N}(\beta)}\right)^{1-s}$ of the scaled partition function via the following relation (see eq.(26) of \cite{FLDR09})
\begin{eqnarray}\label{denmin}
 \ln{\int_{-\infty}^{\infty} p_{\beta}(y)\,e^{ys}\, dy}= \ln
\tilde{M}_{\beta}(1+\frac{s}{\beta}) + \ln  \Gamma(1+\frac{s}{\beta})
\end{eqnarray}
Actually, as  shown in \cite{FLDR09} the freezing scenario implies  that the variable $y$ whose
 probability density is given by $p_{\beta=1}(y)$  is precisely the fluctuating part of
 the height of the global minimum of the random potential which is our main object of interest.
 Note however that the scale $\mathcal{Z}^{(e)}_{N}(\beta)$
 diverges when approaching the critical point $\beta=1$, and that the associated free energy $-\frac{1}{\beta} \log{\mathcal{Z}^{(e)}_{N}(\beta)}$ is self-dual only in the leading order, given by
  $-(\beta+\beta^{-1})\log{N}$. The latter term after freezing at $\beta=1$ yields the leading $2\log{N}$ term in our conjecture Eq.(\ref{predeq}) for the maximum, whereas the logarithmically divergent term $-\frac{1}{\beta}\log{\Gamma(1-\beta^{2})}$  after careful re-interpretation results in the second term $-\frac{3}{2}\log{\log{N}}$, see  \cite{FLDR12} for the detailed explanation of that mechanism. The procedure leaves however a certain arbitrariness in the terms of the order of unity in the mean free energy, hence in the overall shift of the position of the maximum. Let us stress however that apart from such a shift, the shape of the distribution function recovered in the framework of the freezing paradigm is completely fixed by the procedure.

 Our strategy therefore will be to check if self-duality holds for the right-hand side combination in (\ref{denmin})
when we substitute our expression for the moments. Before we proceed, it will be helpful to further expand our expression \eqref{znmoms}. Inserting \eqref{momG} and making use of the identity
\begin{equation}
\frac{1}{G_{\beta}(\beta(s-1)+1/\beta)} = \frac{\Gamma(1+\beta^{2}(s-1))}{G_{\beta}(\beta s+1/\beta)}(2\pi)^{(\beta-1)/2}\beta^{-1/2-\beta^{2}(s-1)}
\end{equation}
shows that (taking into account all prefactors coming from \eqref{mompart}, \eqref{momselb} and \eqref{momG})
\begin{equation}
\begin{split}
&\mathbb{E}(\mathcal{Z}_{N}(\beta)^{1-s}) \sim [\mathcal{Z}^{(e)}_{N}(\beta)]^{1-s}2^{2\beta^{2}+B_{2}s}\pi^{-s}\beta^{\beta^{2}s}\frac{(2\pi)^{(\beta-1)/2}\beta^{-1/2-\beta^{2}(s-1)}}{M^{(G)}_{\beta}(1)}\\
&\times\Gamma(1+\beta^{2}(s-1))\frac{G_{\beta}\left(\beta\left(s+\frac{a_{2}+b_{2}+1}{2}\right)+\frac{2+a_{1}+b_{1}}{2\beta}\right)
G_{\beta}\left(\beta\left(s+\frac{a_{2}+b_{2}}{2}\right)+\frac{2+a_{1}+b_{1}}{2\beta}\right)}
{G_{\beta}\left(\beta(s+a_{2})+\frac{1+a_{1}}{\beta}\right)G_{\beta}\left(\beta(s+b_{2})+\frac{1+b_{1}}{\beta}\right)}\\ \label{momGdual}
&\times \frac{G_{\beta}\left(\beta\left(s+\frac{a_{2}+b_{2}+1}{2}\right)+\frac{3+a_{1}+b_{1}}{2\beta}\right)
G_{\beta}\left(\beta\left(s+\frac{a_{2}+b_{2}}{2}\right)+\frac{3+a_{1}+b_{1}}{2\beta}\right)}
{G_{\beta}\left(\beta(s+1+a_{2}+b_{2})+\frac{2+a_{1}+b_{1}}{\beta}\right)G_{\beta}\left(\beta s+\frac{1}{\beta}\right)}
\end{split}
\end{equation}
A direct inspection makes it clear that the self-duality is only possible if either $a_1=a_2, b_1=b_2$ or
$a_1=b_2, b_1=a_2$.  For the GUE characteristic polynomials, we have $a_{1}=b_{1}=1/2$, $a_{2}=b_{2}=q/2$ so that duality occurs \textit{only if} $q=1$. We therefore have to choose $q=1$ to be able to rely upon the freezing scenario
allowing to interpret the function $p_{\beta=1}(y)$ calculated from its Laplace transform via (\ref{denmin}) as the probability density for the (shifted) global minimum. Using (\ref{momGdual}) with $a_1=a_2=b_1=b_2=\frac{1}{2}$  we get
\begin{equation}
\begin{split}
&\mathbb{E}(\mathcal{Z}^{1-s}_{N}(\beta)) \sim [\mathcal{Z}^{(e)}_{N}(\beta)]^{1-s}2^{(1+\beta^{2})(s-1)}\\
&\times\Gamma(1+\beta^{2}(s-1))\frac{G_{\beta}\left(\beta\left(s+1\right)+\frac{3}{2\beta}\right)}{G_{\beta}\left(\beta(s+1/2)+\frac{3}{2\beta}\right)}\frac{G_{\beta}\left(\beta\left(s+1\right)+\frac{2}{\beta}\right)
G_{\beta}\left(\beta\left(s+\frac{1}{2}\right)+\frac{2}{\beta}\right)}
{G_{\beta}\left(\beta(s+2)+\frac{3}{\beta}\right)G_{\beta}\left(\beta s+\frac{1}{\beta}\right)}c_{\beta}\label{momGfinal}
\end{split}
\end{equation}
where $c_{\beta}$ is a constant determined by the condition $\mathbb{E}(\mathcal{Z}_{N}(\beta)^{1-s})|_{s=1}=1$. Inserting \eqref{momG} into the right-hand side of \eqref{denmin} (which is now manifestly self-dual) leads to the following expression at $\beta=1$:

\begin{align}
&\int_{-\infty}^{\infty} p_{\beta=1}(y)\,e^{ys}\, dy=K^{s}\Gamma(1+s) \tilde{M}_{\beta=1}(1+s) 
\\ &= \frac{1}{C}K^{s}\Gamma^2(1+s) \frac{G(s+7/2)^{2}G(s+3)G(s+4)}{G^2(s+3)G(s+6)G(s+2)}=
\frac{1}{C}K^{s}\frac{\Gamma(1+s)G(s+7/2)^{2}\Gamma(s+3)}{G(s+1)G(s+6)} \label{finalpred}.
\end{align}
where $C = c_{\beta=1}$ and $K$ is a constant which determines the shift in the maximum as discussed below \eqref{denmin}. The value $K=2\pi$ in \eqref{pred} is conjectured from the results of numerical simulations in the next section. The latter formula \eqref{finalpred} constitutes our main analytical result and finally leads to our Conjecture \ref{pred}.

\section{Numerical study of the distribution of the maximum modulus of GUE characteristic polynomials}
The purpose of this Section is to provide a numerical test of Conjecture \ref{pred}.

\subsection{Results}
\begin{figure}
\includegraphics[width=400pt]{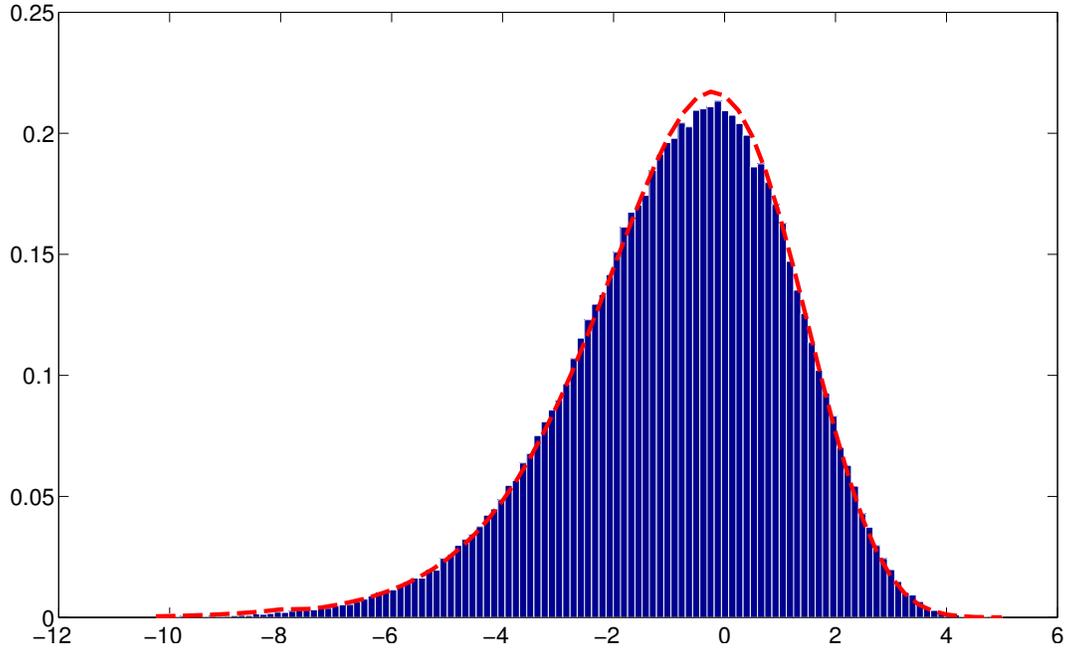}
\caption{The centered and scaled maximum as defined by \eqref{tildep}. The dashed line is the probability density of the random variable $y$ given in Laplace space by \eqref{Laplace}.}
\label{fig:N3000M250kGUE}
\end{figure}
In Figure \ref{fig:N3000M250kGUE} we present a histogram of the recentered and rescaled maximum of the GUE characteristic polynomial, defined by
\begin{equation}
y^{*}_{N} := (2\log(N)-(3/2)\log(\log(N))-M_{N}^{*}+c_{N}^{*})(1+s_{N}^{*}) \label{tildep}
\end{equation}
with $M_{N}^{*}$ defined in (\ref{maxrv}). Here we used the matrix size $N=3000$ and  $250,000$ realizations of the GUE ensemble. The dashed red line is the exact probability density of the random variable $y$ defined via its Laplace transform in \eqref{Laplace}. In \eqref{tildep} we have recentered and scaled by $c_{N}^{*} = 0.216$ and $s_{N}^{*} = 0.188$, presumably a consequence of finite-$N$ effects due to the $o(1)$ terms in \eqref{predeq}. Note that the influence of shift/recentering is already quite small compared with the predicted considerably larger $(3/2)\log(\log(N)) \sim 3.12$ shift. The parameters $c_{N}^{*}$ and $s_{N}^{*}$ were calculated empirically from the mean and variance of $y$ in \eqref{Laplace} according to the formula
\begin{equation}
\begin{split}
s_{N}^{*} &= \sqrt{\mathrm{Var}(y)/\mathrm{Var}(M_{N}^{*})}-1\\
c_{N}^{*} &= \mathbb{E}(y)/(s^{*}_{N}+1)-(2\log(N)-(3/2)\log(\log(N))-M_{N}^{*}) \label{eqcnsn},
\end{split}
\end{equation}
as derived by requiring $\mathbb{E}(y_{N}^{*}) = \mathbb{E}(y)$ and $\mathrm{Var}(y_{N}^{*}) = \mathrm{Var}(y)$. In Table \ref{table:cnsn} we display values of the parameters $c_{N}^{*}$ and $s_{N}^{*}$ for the studied range of sizes $N$, as determined empirically from the mean and variance of the random variable $u$.
\begin{figure}
\includegraphics[width=400pt]{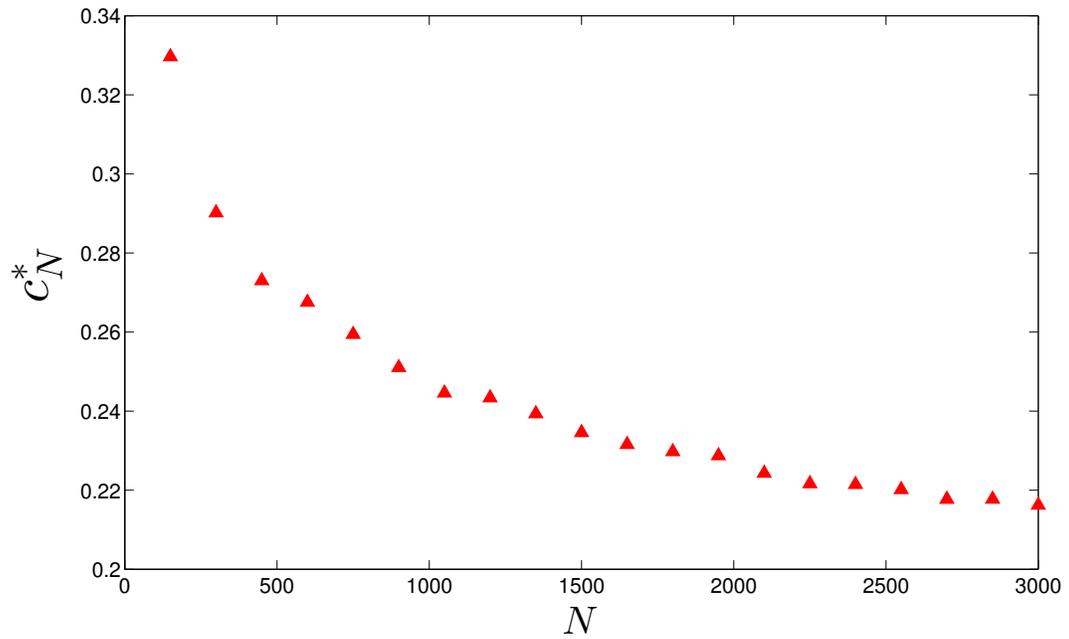}
\caption{Each triangle represents a value of $c_{N}^{*}$ obtained from \eqref{eqcnsn} with $250,000$ realizations.}
\label{fig:cnstar}
\end{figure}
\begin{table}[ht]
\caption{Finite-$N$ corrections for increasing values of $N$ all with $250,000$ realizations}
\centering
\begin{tabular}{c c c c}
\hline\hline
$N$ & $c_{N}^{*}$ & $s_{N}^{*}$ \\ [0.5ex]
\hline
150 & 0.329 & 0.331 \\
600 & 0.267 & 0.248 \\
1050 & 0.244 & 0.224 \\
1500 & 0.234 & 0.212 \\
1950 & 0.228 & 0.202 \\
2400 & 0.221 & 0.195 \\
3000 & 0.216 & 0.188 \\
\hline
\hline
\end{tabular}
\label{table:cnsn}
\end{table}
The observed decay with $N$ is certainly consistent with asymptotic validity of our Conjecture \ref{pred}, though the convergence to the asymptotic results is too slow to make more definite claims. To resolve further decrease of the coefficients $c_{N}^{*}$ and $s_{N}^{*}$ would require much larger matrices and is computationally demanding.

\begin{figure}
\includegraphics[width=400pt]{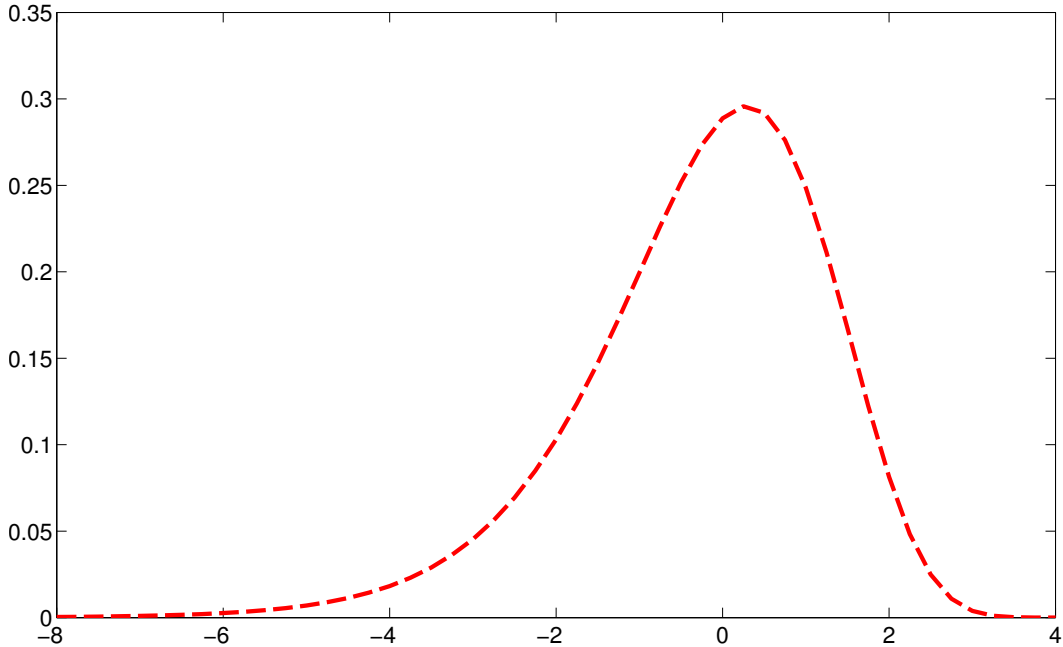}
\caption{The inverse Laplace transform of formula \eqref{uprimelaplace}.}
\label{fig:uprimefig}
\end{figure}

Finally, we provide a numerical validation of the decomposition \eqref{uprime}. In Figure \ref{fig:uprimefig} we plot the inverse Laplace transform of \eqref{uprimelaplace} obtained by a direct numerical evaluation of the integral in the Bromwich inversion formula for the Laplace transform. The positive and normalized curve clearly corresponds to a \textit{bona fide} probability density of some real random variable $y'$.

\subsection{Numerical method}
The numerical evaluation of the maximum value \eqref{maxrv} may be considered quite a non-trivial problem in its own right, for at least two reasons. Firstly, the characteristic polynomial $p_{N}(x)$ having zeros as the eigenvalues of $\mathcal{H}$, displays $O(N)$ oscillations in the spectral interval $[-1,1]$ with hugely varying peaks heights. This produces considerable clusterings of `near-maxima' which may confuse any naive attempt to find the true maximum value. Secondly, the slow changing nature of the correction terms in Conjecture \ref{pred}, of order $\log(N)$ and $\log\log(N)$)  respectively, require one to go to somewhat large matrices to resolve reasonable asymptotic behaviour. The problem is further compounded by the numerical instability of calculating determinants of such matrices.

Our solution to these problems heavily relies on a sparse realization of GUE matrices $H$ originally due to Trotter \cite{T84} (see also Dumitriu and Edelman \cite{DE2002}). He discovered that the eigenvalues of GUE matrices $H$ have the same joint probability density as those of the following real symmetric tri-diagonal matrix:
\begin{equation}
\mathcal{H} = \frac{1}{2\sqrt{2N}} \begin{pmatrix}
\mathcal{N}(0,2) & \chi_{2}    &               &         & & \\
\chi_{2}    & \mathcal{N}(0,2) & \chi_{4} &         & &\\
                 &  \ddots          & \ddots        & \ddots\\
                 &                  & \chi_{2(N-2)}      &\mathcal{N}(0,2) & \chi_{2(N-1)} \\
                 &                  &               & \chi_{2(N-1)}         & \mathcal{N}(0,2)\\

\end{pmatrix}
\end{equation}
where $\mathcal{N}(0,2)$ is a normal random variable with mean $0$ and variance $2$. The sub-diagonal is composed of random variables $\chi_{2n}$ having the same density as $\sqrt{\chi^{2}_{2n}}$ where $\chi^{2}_{2n}$ is a $\chi$-square random variable with $2n$ degrees of freedom. To compute the maximum value of $p_{N}(x) = \det(xI-H) = \det(xI-\mathcal{H})$, we begin by exploiting the known asymptotic behaviour
\begin{equation}\label{asympmean}
2\mathbb{E}\log|p_{N}(x)| = N(2x^{2}-1-2\log(2))+o(1)
\end{equation}
so that
\begin{equation}
\begin{split}
f_{N}(x) &:= 2\log|p_{N}(x)|-2\mathbb{E}\log|p_{N}(x)|  \sim 2\log|\det(e^{-(x^{2}-1/2-\log(2))}(xI-\mathcal{H}))|
\end{split}
\end{equation}
Further progress is now possible thanks to the fact that determinants of tri-diagonal matrices satisfy a linear recurrence relation. Furthermore, by an appropriate rescaling, the recursion computes determinants of all leading principal minors simultaneously, thus computing $f_{j}(x)$ \textit{for all} $j=1,\ldots,N$ in linear time.

Now to find the maximum, we define a mesh $\mathcal{M} = \{-1+n/\Delta : n=0,\ldots,2\Delta\}$ with $\Delta \sim 2N$ and evaluate $f_{N}(x)$ at each of the points in $\mathcal{M}$. At those points where $f_{N}(x)$ is maximal the Matlab function `fminbnd' is invoked to converge onto the global maximum. Figure \ref{fig:N3000gueres} illustrates the complexity of the problem. Our algorithm is sufficiently precise to distinguish the true maximum (located at $x \approx -0.3$ in red) from other possible candidates, \textit{e.g.} $x \approx -0.7$ as well as the thousands of other local maxima.

\end{document}